\documentclass[aps,prl,preprintnumbers,twocolumn,superscriptaddress,showpacs,floatfix]{revtex4}
\usepackage{epsfig}
\usepackage{bm}
\usepackage{amssymb}
\usepackage{amsmath}
\usepackage{color}
\usepackage{subfigure}
\usepackage{hyperref}
\usepackage{dcolumn}

\newcommand{\be}{\begin{equation}}
\newcommand{\ee}{\end{equation}}
\newcommand{\bea}{\begin{eqnarray}}
\newcommand{\eea}{\end{eqnarray}}

\newcommand{\bk}{{\bm{k}}}
\newcommand{\bp}{{\bm{p}}}
\newcommand{\bx}{{\bm{x}}}
\newcommand{\eq}{{\,=\,}}
\newcommand{\uh}[1]{{\color{black} #1}} % change {red} --> {black} when done

\begin{document}

\title{Analytic solution of the Boltzmann equation in an expanding system}
\date{\today }
\author{D.~Bazow}
\affiliation{Department of Physics, The Ohio State University, Columbus, OH 43210, USA}
\author{G.~S.~Denicol}
\affiliation{Department of Physics, McGill University, 3600 University Street, Montreal,
QC, H3A 2T8, Canada}
\affiliation{Physics Department, Brookhaven National Lab, Building 510A, Upton, NY, 11973, USA}
\author{U.~Heinz}
\affiliation{Department of Physics, The Ohio State University, Columbus, OH 43210, USA}
\author{M.~Martinez}
\affiliation{Department of Physics, The Ohio State University, Columbus, OH 43210, USA}
\author{J.~Noronha}
\affiliation{Instituto de F\'{\i}sica, Universidade de S\~{a}o Paulo, C.P. 66318,
05315-970 S\~{a}o Paulo, SP, Brazil}
\affiliation{Department of Physics, Columbia University, 538 West 120th Street, New York,
NY 10027, USA}

\begin{abstract}
\uh{For a massless gas with constant cross section in a homogeneous, isotropically expanding spacetime we reformulate the relativistic Boltzmann equation as a set of non-linear coupled moment equations. For a particular initial condition this set can be solved exactly, yielding the first analytical solution of the Boltzmann equation for an expanding system. The non-equilibrium behavior of this relativistic gas can be mapped onto that of a homogeneous, static non-relativistic gas of Maxwell molecules.} 
\end{abstract}



\maketitle

%Alphabetical order ...

\noindent \textsl{1. Introduction.} The relativistic Boltzmann equation plays a prominent role in understanding the complex non-equilibrium dynamics displayed by dilute relativistic gases. It has applications in many areas of physics including, e.g., theoretical description of the quark-gluon plasma \cite{Heinz:1984yq, Bass:1998ca, Arnold:2000dr, Molnar:2001ux,Xu:2004mz,Denicol:2012cn}, neutrino transport in supernovae \cite{Liebendoerfer:2003es, Janka:2006fh}, and structure formation in cosmology \cite{Ma:1995ey, dodelson, Weinberg:2008zzc}. While analytical solutions of the Boltzmann equation have been thoroughly studied for homogeneous systems \cite{Bobylev1976,krookwu,ernst}, solutions for an expanding system remain to be found even in the non-relativistic regime. 

In the relativistic regime, progress \uh{in this direction was made recently} \cite{Florkowski:2013lza, Florkowski:2013lya, Denicol:2014mca, Denicol:2014xca, Denicol:2014tha, Nopoush:2014qba, Noronha:2015jia, Hatta:2015kia} using the Anderson-Witting equation \cite{aw}, an approximation of the Boltzmann equation that relies on the relaxation time approximation \cite{Bhatnagar:1954zz}. In this scheme, the nonlinear collision kernel of the Boltzmann equation is replaced by a \uh{linearized} version that qualitatively describes the relaxation of the system to equilibrium on a microscopic time scale. These analyses served to improve our understanding of the domain of applicability of a number of extended hydrodynamic theories used in the description of rapidly expanding plasmas \cite{Florkowski:2013lza, Florkowski:2013lya, Denicol:2014mca, Denicol:2014xca, Denicol:2014tha, Nopoush:2014qba}, with focus on the description of ultrarelativistic heavy-ion collisions. 

However, a complete description of dilute gases can only be achieved by solving the \uh{full} Boltzmann equation. While this can be done numerically, \uh{simple yet physically motivated analytical solutions of the Boltzmann equation, if available, can lead to valuable insights} into non-equilibrium phenomena. In this Letter we take a step in this direction and derive the first analytical solution of the \uh{full} Boltzmann equation for an expanding dilute gas. This is done using the method of moments \cite{grad,Denicol:2012cn} to calculate the nonlinear collision term of the relativistic Boltzmann equation for a massless gas with constant cross section in a homogeneous and isotropically expanding spacetime. The derived solution describes the expansion-driven deviation of the \uh{particles'} momentum distribution from local thermal equilibrium as a function of the system's expansion rate. We note that the non-expanding limit of this solution, \uh{which describes the homogeneous relaxation of a {\em relativistic} gas towards equilibrium, is also new.}

\noindent \textsl{2. Boltzmann equation.} We consider a homogeneous and isotropically expanding system of massless particles in a Friedmann-Lema\^itre-Robertson-Walker (FLRW) metric \cite{weinberg,baumann} (the most general homogeneous and
isotropic metric in flat space) 
\begin{equation}
ds^{2}=dt^{2}-a^{2}(t)\left( dx^{2}+dy^{2}+dz^{2}\right) \,.  
\label{metric}
\end{equation}%
We note that in this paper we make the further restriction of zero global curvature. \uh{For} a gas of massless particles, the energy density and the pressure are related as $\varepsilon\eq3p$, and here the scale factor $a(t){\,>\,}0$ \cite{baumann} is a free function that may be fixed
by additional physics assumptions. For the FLRW metric the nonzero Christoffel symbols are $\Gamma_{0j}^{i}\eq\delta _{j}^{i}\,H(t)$ and $\Gamma _{ij}^{0}\eq{a(t)}\dot{a}(t)\,\delta _{ij}$, where $i,j$ denote spatial indices, $H(t)\equiv \dot{a}(t)/a(t)$ is the Hubble parameter, and $\sqrt{-g}=a^{3}(t)$, with $g$ being the determinant of the metric in \eqref{metric}. Even though the fluid flow of this system is locally static, $u^{\mu }\eq\left(1,0,0,0\right) $, the expanding FLRW geometry induces a nonzero fluid expansion rate $\theta (t)\equiv \partial _{\mu }(\sqrt{-g}\,u^{\mu })/\sqrt{-g}=3H(t)$.

The dynamics of the single-particle distribution function, $f_\bk(x)$, is given by the relativistic Boltzmann equation in curved space \cite{debbasch,book,bernstein,Denicol:2014xca,Denicol:2014tha}
\begin{equation}
k^{\mu }\partial _{\mu }f_\bk+\Gamma _{\mu i}^{\lambda }k_{\lambda} k^{\mu } 
\frac{\partial f_\bk}{\partial k_{i}}=\mathcal{C}[f].
\label{Boltzmann1}
\end{equation}
Since the FLRW geometry is based on the assumptions of spatial homogeneity and isotropy, $f_\bk$ cannot depend on spatial position $\bx$ and must be locally isotropic in momentum, depending only on $u{\cdot}k\eq{k}^0$ where for massless particles $k^0\eq{k}/a(t)$ with $k\eq|\bk|$. We therefore write $f_\bk(x)=f_k(t)$ from here on.

The symmetries of the FLRW metric strongly constrain the form of the energy-momentum tensor $T^{\mu \nu }$ and particle 4-current $N^{\mu }$ of the matter. Due to local momentum isotropy the viscous shear-stress tensor and particle diffusion current vanish exactly. Further, the bulk viscous pressure is zero for massless particles. Therefore, the conserved currents always take their equilibrium form, and the time
evolution of the energy and particle densities, $\varepsilon$ and $n$, is fully determined by the conservation laws
\begin{equation}
\partial _{t}n+3nH(t)=0,\qquad \partial _{t}\varepsilon +4\varepsilon H(t)=0.
\label{eqsenergydensity}
\end{equation}
With initial condition $a(t_{0})\eq1$, they are solved by $n(t)\eq{n(t_{0})}/a^{3}(t)$ and $\varepsilon (t)\eq\varepsilon (t_{0})/a^{4}(t)$.

Even though the conserved currents $T^{\mu\nu}\eq\varepsilon u^\mu u^\nu-pg^{\mu\nu}$ and $N^\mu{\eq}n u^\mu$ take the same form as in local thermal equilibrium, the system itself does not have to be in equilibrium. In fact, the local momentum distribution $f_k(t)$ is driven away from its local equilibrium form by an amount proportional to the expansion rate $\theta(t)\eq3H(t)$ of the FLRW geometry. However, this process happens without disturbing the spatial homogeneity and isotropy of the system and, consequently, there are no out-of-equilibrium corrections to the conserved currents. A similar behavior was found in \cite{Noronha:2015jia,Hatta:2015kia} in the context of the relaxation time approximation.

\uh{The contributions from the Christoffel symbols in Eq.~(\ref{Boltzmann1}) cancel exactly, and the equation} reduces to 
\begin{equation}
k^{0}\partial _{t}f_k=\mathcal{C}[f].
\end{equation}
Ignoring quantum statistics, the collision term \uh{reads}
\begin{equation}
\mathcal{C}[f]=\frac{1}{2}\int_{k'pp'} \!\! W_{\bk{\bk'}\to\bp{\bp'}} \left( f_pf_{p'}{-}f_k f_{k'}\right),
\label{collision1}
\end{equation}
where $\int_k{\,\equiv\,}\int d^3k/\left[(2\pi)^{3}\sqrt{-g}\,k^{0}\right]$ and $W_{\bk{\bk'}\to\bp{\bp'}}$ is the transition rate. To make progress on evaluating (\ref{collision1}) we make the simplifying assumption of isotropic scattering with energy-independent total cross section $\sigma$. Then $W_{\bk{\bk'}\to\bp{\bp'}}$ takes the form \cite{book,bernstein,debbasch} 
\begin{equation}
W_{\bk{\bk'}\to\bp{\bp'}} = (2\pi)^{5}\sqrt{{-}g}\,\sigma\,s\,\delta^4(k{+}k'{-}p{-}p'),
\label{definerate}
\end{equation}
where $s\eq(k^\mu{+}{k'}^\mu)(k_\mu+k'_\mu)$. Thus, the Boltzmann equation becomes
\begin{equation}
k^0\partial_t f_k\eq\frac{(2\pi)^{5}}{2} \sqrt{{-}g}\, \sigma\!\! \int_{k'pp'} \!\!\!\!\!\!
s\,\delta^4(k{+}k'{-}p{-}p')(f_p f_{p'}{-}f_k f_{k'}).  \quad  
\label{Boltzmanneq2}
\end{equation}
Note that even for the highly symmetric case considered here the relativistic Boltzmann equation is still a non\-linear integro-differential equation for $f_k$ .

\noindent \textsl{3. Moment equations.} We proceed to solve the Boltzmann equation using the method of moments \cite{Denicol:2012cn}. Due to local momentum isotropy, the distribution function $f_k$ can
be fully described by scalar moments only \cite{fn1}:
\begin{eqnarray}
\rho _{m}(t) &=& \int_\bk (u{\cdot}k)^{m+1}\,f_k(t) = \int_\bk (k^0)^{m+1}\,f_k(t)
\\\nonumber
&=& \frac{1}{2\pi^{2}}\frac{1}{a^{m+3}(t)}\int_{0}^{\infty} dk\,k^{m+2}\,f_k(t)
\quad (m\in\mathbb{N}_0).\quad
\label{definerhomoments}
\end{eqnarray}
The time dependence of the two lowest moments, the number density $n(t){\,\equiv\,}\rho _0(t)$ and the energy density $ \varepsilon(t){\,\equiv\,}\rho_1(t)$, is given by the conservation laws already discussed. For a classical gas of massless particles, they provide the time-dependence of the temperature $T$ and fugacity $\lambda\eq\exp(\mu/T)$ (where $\mu$ is the chemical potential) via the matching conditions \cite{Denicol:2012cn} $T\eq\varepsilon/(3n)$ and $\nu_d\lambda{\eq}n\pi^{2}/T^{3}$ where $\nu_d$ is the number of massless degrees of freedom (including statistical degeneracy factors). 
In the following we set $\nu_d\eq1$ for simplicity. 

From the matching conditions it follows that $T(t){\eq}T(t_0)/a(t)$ and $\mu(t)\eq\mu(t_0)/a(t)$ such that the fugacity $\lambda$ is time-independent. $T(t)$ and $\lambda$ define the local equilibrium distribution function $f_k^{\mathrm{eq}}(t)=\lambda\exp(-u{\cdot}k/T(t))$ and the equilibrium scalar moments
\begin{equation}
\rho _{m}^{\mathrm{eq}}(t)=\frac{(m{+}2)!}{2\pi^{2}}\lambda\,T^{m{+}3}(t).
\label{eqf}
\end{equation}
To express the Boltzmann equation in terms of the scalar moments $\rho_m$ we multiply \eqref{Boltzmanneq2} by $(u{\cdot}k)^m $ and integrate over $k$. This results in
\begin{equation}
\label{momentevolution}
\partial _{t}\rho _{m}(t)+(3{+}m)H(t)\rho _{m}(t)=\mathcal{C}_{\mathrm{gain}}^{(m)}(t)
-\mathcal{C}_{\mathrm{loss}}^{(m)}(t),
\end{equation}
where the gain and loss terms are defined by
\begin{eqnarray}
&&\!\!\!\!\!\!
\mathcal{C}_{\mathrm{gain}}^{(m)}(t) =
\frac{(2\pi)^{5}}{2} \sqrt{{-}g}\, \sigma\!\! \int_{kk'pp'} \!\!\!\!\!\!\!\!
s\,(u{\cdot}p)^m\delta^4(k{+}k'{-}p{-}p') f_k f_{k'}, 
\label{gain1} 
\nonumber\\
&&\!\!\!\!\!\!
\mathcal{C}_{\mathrm{loss}}^{(m)}(t) =
\frac{(2\pi)^{5}}{2} \sqrt{{-}g}\, \sigma\!\! \int_{kk'pp'} \!\!\!\!\!\!\!\!
s\,(u{\cdot}k)^m\delta^4(k{+}k'{-}p{-}p') f_k f_{k'}.
\notag\\
\end{eqnarray}
Here the gain term was simplified by using the symmetry of the transition rate under interchange of the incoming and outgoing momenta. With
\begin{equation}
\sqrt{-g} \int_{pp'} \delta^4(k{+}k'{-}p{-}p')=1/(2\pi)^5,
\end{equation}
the loss term reduces straightforwardly to \cite{Denicol:2012cn}
\begin{equation}
\mathcal{C}_{\mathrm{loss}}^{(m)}(t)=\sigma \rho _{m}(t)n(t).
\end{equation}
However, the gain term is more involved. One first writes

\begin{equation}
\mathcal{C}_{\mathrm{gain}}^{(m)}(t)=\frac{\sigma}{2} \int_{kk'} s\,f_k f_{k'}\,\mathcal{P}_{m},
\label{definenewgain}
\end{equation}
where the inner kernel is
\begin{equation}
\mathcal{P}_{m}\equiv (2\pi)^5\,\sqrt{-g} \int_{pp'} (u{\cdot}p)^m\,\delta^4(k{+}k'{-}p{-}p').
\label{definePn}
\end{equation}

This quantity is a scalar and can be calculated in any frame. In the center of momentum frame, the 4-dimensional delta function can be integrated and the remaining angular integrals can be performed analytically \cite{longpaper}. The final result can be expressed as follows
\begin{equation}
\label{Pm}
\mathcal{P}_{m} = \frac{m!}{[2a(t)]^m} \sum_{j\,\mathrm{odd}}^{m+1} 
\frac{(k{+}k')^{m{+}1{-}j}}{(m{+}1{-}j)!}\, \frac{\left\vert \bk{+}\bk'\right\vert^{j-1}}{j!}\,.  
\end{equation}
Plugging this back into \eqref{definenewgain} and performing some
algebraic manipulations \cite{longpaper} one obtains

\begin{equation}
\mathcal{C}_{\mathrm{gain}}^{(m)}(t) = 2\,(m{+}2)\,m!\,\sigma \sum_{j=0}^{m}
\frac{\rho_j(t)}{(j{+}2)!}\,\frac{\rho_{m-j}(t)}{(m{-}j{+}2)!}.  
\label{gainmoneyshot}
\end{equation}
Using these results in Eq.~(\ref{momentevolution}) we derive the following exact evolution equation for
the moments:
\begin{eqnarray}
&&\partial_t\rho _{m}(t) +[(3{+}m)H(t) + \sigma n(t)]\rho_m(t) 
\nonumber\\
&&=2\,(m{+}2)\,m!\,\sigma \sum_{j=0}^{m}\frac{\rho_j(t)}{(j{+}2)!}\,\frac{\rho_{m-j}(t)}{(m{-}j{+}2)!}.
\label{equationrhon}
\end{eqnarray}
This infinite set of coupled nonlinear differential equations for the scalar moments $\rho _m(t)$ is equivalent to the original relativistic integro-differential Boltzmann equation. For $m\eq0$ and $m\eq1$ Eq.~(\ref{equationrhon}) reduces to Eqs.~\eqref{eqsenergydensity} for the particle and energy densities. \uh{Higher-order moments are more sensitive to the distribution function at higher momentum.} We can further simplify the moment equations by introducing the scaled moments $M_m(t){\,\equiv\,}\rho_m(t)/\rho_m^\mathrm{eq}(t)$ \cite{fn2} and the scaled time $\hat{t}{\eq}t/\ell_0$ where $\ell_0\eq1/(\sigma\,n(t_0))$ is the (constant) mean free path at time $t_0$. This yields the surprisingly simple evolution equations
\begin{equation}
\!\!\!
a^{3} (\hat{t}\,)\,\frac{\partial M_{m}(\hat{t})}{\partial \hat{t}} + M_{m}(\hat{t})
= \frac{1}{m{+}1}\sum_{j=0}^{m}M_{j}(\hat{t})M_{m-j}(\hat{t}).
\label{foda1}
\end{equation}
Equation~\eqref{foda1} is the main result of this Letter. While the nonlinear coupling between different moments was expected from the nonlinearity of the collision kernel, the same cannot be said about another key feature of Eq.~\eqref{foda1}: it can be solved recursively, i.e. the solution of the evolution equation for $M_n$ requires only previously solved moments  $M_k(t)$ of lower order $k{\,<\,}n$. This property depends on our choice of an energy-independent cross section; it is essential for being able to solve \eqref{foda1} analytically.

All information about the expansion appears in the factor $a^3(\hat{t}\,)$ multiplying the time derivative in \eqref{foda1}. \uh{Whether or not} local equilibrium can be achieved \uh{thus} depends on the state of expansion of the system. For example, when $a(\hat{t}\,){\,\sim\,}\hat{t}\,^{1/2}$ the mean free path $\ell(\hat{t})\eq1/(\sigma n(\hat{t}))$ increases faster than the expansion rate, $\lim_{\hat{t}\to \infty} \ell(\hat{t})\theta(\hat{t}){\,\to\,}\infty$, and local equilibrium cannot be reached \uh{even} at asymptotically large times. Moreover, if the initial $f_k(\hat{t}_{0})$ is positive definite, $M_{n}(\hat{t}_{0}){\,>\,}0$, and Eq.~\eqref{foda1} then implies that all moments remain positive throughout the evolution, translating into positivity for $f_k$ for all momenta at all times.

Equation \eqref{foda1} closely resembles the Bobylev-Krook-Wu (BKW) equation derived almost four decades ago \cite{Bobylev1976, krookwu} in a famous study about homogeneous and isotropic solutions of the \emph{non-relativistic} Boltzmann equation (for a review see \cite{ernst}) for Maxwell molecules. In fact, by defining the time variable $\tau =\int_{\hat{t}_{0}}^{\hat{t}}dt'/a^3(t')$ to take into account the expansion \cite{fn3}, our equation \eqref{foda1} for the moments becomes identical to the BKW equations \cite{krookwu}:
\begin{equation}
\label{BKW}
\partial_\tau M_{m}(\tau)+M_{m}(\tau)=\frac{1}{m{+}1}\sum_{j=0}^{m}M_{j}(\tau)M_{m-j}(\tau).
\end{equation}
This indicates that even though the underlying symmetries of these physical systems are quite different (BKW's are based on Galilean invariance with static conditions while ours are embedded in an expanding system), these systems are actually equivalent from a dynamical perspective and evolve towards equilibrium in a universal manner. We note that our equations also reduce to BKW's when $a(\hat{t}){\,\equiv\,}1$, i.e. for a non-expanding metric (though special relativistic effects are still fully taken into account).

\begin{figure}[th]
\includegraphics[width=0.9\linewidth]{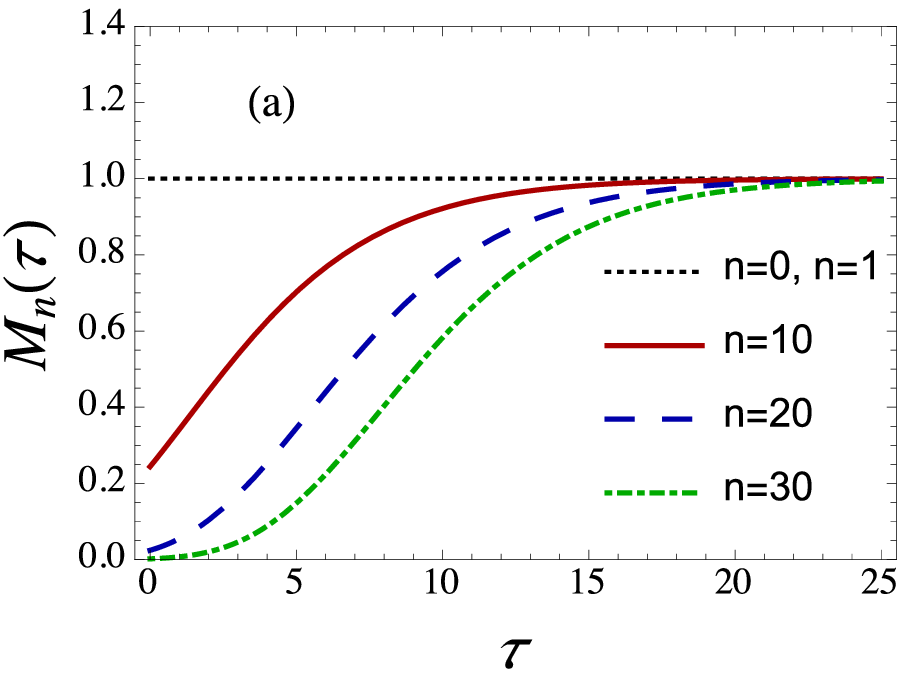}\\
\includegraphics[width=0.9\linewidth]{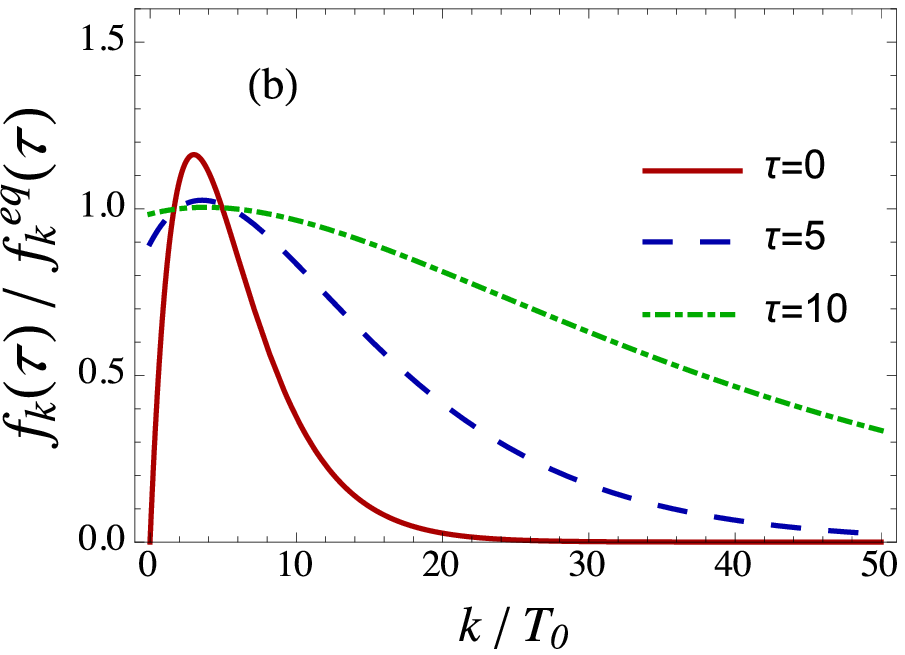}
\caption{(Color online) Time evolution of the moments $M_n$ (a) and of the ratio between the
out-of-equilibrium solution \eqref{fullanalboltzmann} and the equilibrium distribution (b).}
\label{fig1}
\end{figure}

\noindent \textsl{4. Analytical solution of the moment equations.} Given the close relation of our Eq.~(\ref{foda1}) with the BKW equation (\ref{BKW}) it is not surprising that it admits an exact analytic solution of the Krook and Wu type:
\begin{equation}
M_{m}(\tau )=\mathcal{K}(\tau )^{m-1}\bigl[ m-(m{-}1)\mathcal{K}(\tau )\bigr] \quad (m\geq0),
\label{KWmoments}
\end{equation}
where $\mathcal{K}(\tau )=1-\frac{1}{4}\exp(-\tau/6)$. The time evolution of the moments is shown in Fig.~\ref{fig1} (a). One sees that low-order moments equilibrate more quickly than the higher-order ones that are needed to describe the high-momentum non-equilibrium tail of the distribution function.

Given the exact form (\ref{KWmoments}) of the moments $M_m(\tau)$, the distribution function can be reconstructed as follows. One defines the function $\mathcal{F}(\tau,u{\cdot}k)\eq(u{\cdot}k)^2\,\theta (u{\cdot}k)\,f_k$ where $u{\cdot}k{\eq}k/a(\tau)$, expands its Fourier transform with respect to $k$ in terms of the moments $M_m(\tau)$, and then transforms it back. This yields the exact result \cite{longpaper}
\begin{eqnarray}
\label{fullanalboltzmann}
f_k(\tau) &=&\lambda\,\exp \left( -\frac{u{\cdot}k}{\mathcal{K}(\tau)T(\tau)}\right) 
\\ \nonumber &\times & \left[ \frac{4\mathcal{K}(\tau){-}3}{\mathcal{K}^{4}(\tau )}
+\frac{u{\cdot}k}{T(\tau)}\left( \frac{1{-}\mathcal{K}(\tau)}{\mathcal{K}^{5}(\tau)}\right) \right]. 
\notag
\end{eqnarray}

To the best of our knowledge, this is the first \uh{analytic solution of the full} Boltzmann equation for an expanding interacting gas. At the initial time $\hat{t}_{0}$ (corresponding to $\tau\eq0$ and $T(\tau){\eq}T_0$) one finds the initial condition $f_k(0) = \frac{256}{243}\,(k/T_0)\,\lambda\,\exp[-4k/(3T_0)] >0$. In Fig.~\ref{fig1} (b) we plot the analytical solution for the ratio $f_k(\tau)/f_{k}^{eq}(\tau)$ as a function of $u{\cdot}k/T(\tau){\eq}k/T_0$ for $\tau=0$, $5$, and $10$. At $\tau \eq0$ the high momentum tail is largely underpopulated whereas momentum modes in the range $k/T_0\sim 1.6-5$ are over-occupied relative to local equilibrium. As time evolves, high momentum modes are populated at the expense of the over-populated moderate momentum region, in a process resembling an energy cascade. Note, however, that when $a(\hat{t}) \sim \hat{t}^{1/2}$ one finds that $\lim_{\hat{t}\to \infty}\tau(\hat{t})$ is finite and \eqref{fullanalboltzmann} never assumes the equilibrium form.

%%%%%%%%%%%%%%% Fig. 1 %%%%%%%%%%%%%%%%%%%%%%%%%
%\begin{figure}[th]
%\subfigure[]{\includegraphics[width=0.40\textwidth]{plotMn_opt.pdf}
%\label{fig1a}
%} 
%\subfigure[]{\includegraphics[width=0.40\textwidth]{plotf_opt.pdf}
%\label{fig1b}
%}
%\caption{(Color online) Time evolution of the moments $M_n$ (a) and of the ratio between the
%out-of-equilibrium solution \eqref{fullanalboltzmann} and the equilibrium distribution (b).}
%\label{fig1}
%\end{figure}
%%%%%%%%%%%%%%%%%%%%%%%%%%%%%%%%%%%%%%%%%%%%%

%%%%%%%%%%%%%%% Fig. 1 %%%%%%%%%%%%%%%%%%%%%%%%%
%%%%%%%%%%%%%%%%%%%%%%%%%%%%%%%%%%%%%%%%%%%%%

The moment equations \eqref{foda1} can \uh{also} be studied numerically for a variety of initial conditions for which no analytic solution is known. Here we briefly address the question whether the direct energy cascade seen in Fig.~\ref{fig1} is related to self-similar behavior characteristic of turbulence \cite{Berges:2015kfa}. While a detailed study of self-similarity in this system is beyond the scope of this \uh{short} paper, we note here that expressing the Boltzmann equation in terms of its moments may be quite useful in this context. For instance, one can show that non-thermal fixed points corresponding to exact self-similar solutions of the kind
\begin{equation}
f_k(\hat{t}) = a^\gamma(\hat{t})\, f_S(a^\beta(\hat{t})\,u{\cdot}k),
\label{scaling}
\end{equation}
where $\gamma$ and $\beta$ are the scaling exponents and $f_S$ is the fixed point distribution \cite{Berges:2015kfa}, are not compatible with Eq.~\eqref{foda1}. In fact, using Eq.~\eqref{eqsenergydensity} one can see that the scaling exponents are necessarily $\gamma\eq0$ and $\beta\eq1$, and that the Ansatz (\ref{scaling}) for the distribution function (if assumed to be valid for all momenta)
leads to moments
\begin{equation}
\rho_m^{S}(\hat{t}) = \frac{1}{2\pi^2}\frac{c_m^S}{a^{m+3}(\hat{t}\,)}
\end{equation}
(where $c_m^S\eq\int_0^\infty d\xi\, \xi^{m+2}\,f_S(\xi)$) \uh{that} have the same time dependence as the equilibrium moments. This implies that the corresponding normalized moments $M_m^S = \rho_m^S/\rho_m^{eq}$ are time independent. Since the conservation laws require that $M_0^S=M_1^S=1$, one can then use \eqref{foda1} to show that $M_m^S\to1$ for all $m$. This shows that there are no other true fixed points of the dynamics besides local thermal equilibrium. 

\noindent \textsl{6. Conclusions.} We derived from the relativistic Boltzmann equation a nonlinear set of \uh{coupled} moment equations \uh{for} a massless gas with constant cross section in a homogeneous, isotropically expanding spacetime. For a particular initial condition, we \uh{found} that the moment equations can be solved exactly, \uh{thereby obtaining} the first analytical solution of the Boltzmann equation \uh{with full non-linear collision term} for an expanding system. 
     
\uh{The normalized moments} of this \uh{expanding} relativistic gas can be directly mapped onto the corresponding moments for a homogeneous, \uh{static} non-relativistic gas of Maxwell molecules. This happens even though the distribution functions of these systems are not the same. \uh{This} nontrivial correspondence suggests that exact self-similar \uh{non-equilibrium} solutions may not exist even in rapidly expanding systems. 

The study performed here can be extended along several directions. One may consider different types of cross sections, or include nonzero particle masses to investigate bulk viscous effects. If possible, the generalization of the method presented here to expanding systems with different symmetries \cite{Florkowski:2013lza,Denicol:2014xca,Gelis:2013rba,Berges:2013fga,Kurkela:2015qoa} that are relevant for the study of
the quark-gluon plasma formed in heavy ion collisions would be particularly desirable.

%%%%%%%%%%%%%%%%%%%%%%%%%%%%%%%%%%%%

\noindent \textsl{Acknowledgements.} GSD thanks S.~Schlichting and A.~Dumitru for enlightening discussions. MM thanks Y.~Mehtar-Tani and S.~Ozonder for useful discussions. JN thanks the Physics Department at Columbia University for its hospitality and Conselho Nacional de Desenvolvimento Cient\'{\i}fico e Tecnol\'{o}gico (CNPq) and Funda\c{c}\~{a}o de Amparo \`{a} Pesquisa do Estado de S\~{a}o Paulo (FAPESP) for financial support. UH and MM express their gratitude to the Institute for Nuclear Theory for hospitality during the final stages of this work. GSD thanks the Kavli Institute of Theoretical Physics China (KITPC) for hospitality during the initial stages of this work. GSD is currently supported under DOE Contract No. DE-SC0012704 and acknowledges previous support of a Banting fellowship provided by the Natural Sciences and Engineering Research Council of Canada. DB, UH, and MM are supported by the U.S. Department of Energy, Office of Science, Office of Nuclear Physics under Award \rm{DE-SC0004286}. UH, MM and JN gratefully acknowledge support through a bilateral travel grant from FAPESP and the Ohio State University.

%%%%%%%%%%%%%%%%%%%%%%%%%%%

%%%%%%%%%%%%%%%%%%%%%%%%%%%%


\begin{thebibliography}{99}
\bibitem{fn2}
Eqs.~\eqref{eqsenergydensity} and the matching conditions imply that $M_{0}(\hat{t}){\eq}M_{1}(\hat{t})\eq1$ throughout the evolution.
\bibitem{fn3}
This is equivalent to scaling the time $t$ by the time-dependent mean free path $\ell(t)$ instead of the constant $\ell_0$ used above Eq.~(\ref{foda1}).
\end{thebibliography}
\end{document}